\documentclass[conference]{IEEEtran}

\usepackage{pgf,tikz}
\usetikzlibrary{arrows}
\usepackage{research5IEEE}

\usepackage{dsfont}
\usepackage{accents}
\usepackage{pifont}

\usepackage{amsmath}
\usepackage{amssymb}
\usepackage{color}
\usepackage{bbm}
\allowdisplaybreaks
\usepackage{cite}
\usepackage{url,epstopdf}
\usepackage{todonotes}
\newtheorem{theorem}{Theorem}
\newtheorem{definition}{Definition}
\newtheorem{lemma}{Lemma}

\newtheorem{remark}{Remark}

\ifCLASSINFOpdf
\else
\fi

\title{Distributed Sequential Hypothesis Testing With Zero-Rate Compression}

\author{\IEEEauthorblockN{Sadaf Salehkalaibar\IEEEauthorrefmark{1} and Vincent Y. F. Tan\IEEEauthorrefmark{2}}
	\IEEEauthorblockA{\small\IEEEauthorrefmark{1} ECE Department, College of Engineering, University of Tehran, Tehran, Iran, Email:
		\url{s.saleh@ut.ac.ir}}
	\IEEEauthorblockA{\small\IEEEauthorrefmark{2} Department of ECE, National University of Singapore, Email:
		\url{vtan@nus.edu.sg}}}

\begin{document}
\maketitle
\begin{abstract}
In this paper, we consider sequential testing over a single-sensor, a single-decision center setup. At each time instant $t$, the sensor gets $k$ samples $(k>0)$ and describes the observed sequence until time $t$ to the decision center over a zero-rate noiseless link. The decision center sends a single bit of feedback to the sensor to request for more samples for compression/testing or to stop the transmission. We have characterized the optimal exponent of type-II error probability under the constraint that type-I error probability does not exceed a given threshold $\epsilon\in (0,1)$ and also when the expectation of the number of requests from decision center is smaller than $n$ which tends to infinity. Interestingly, the optimal exponent coincides with that for fixed-length hypothesis testing with zero-rate communication constraints.  
\end{abstract}


%
\IEEEpeerreviewmaketitle

\section{Introduction}

Hypothesis testing aims at detecting the distribution of sources observed at sensors in networks such as the
Internet of Things (IoT). In a distributed setting, the sensors observe source sequences  and send compressed versions of these observations over the network to a decision center where the underlying distribution of the sources should be detected.

The simplest case of a hypothesis testing setup consists of a sensor which itself should decide on the hypothesis. Assume that there are two hypotheses $\mathcal{H}=0$ (null hypothesis) and $\mathcal{H}=1$ (alternative hypothesis). Under the null and alternative hypotheses, the source $X$ is distributed according to given pmfs $P_X$ and $Q_X$, respectively. The performance of this system is characterized by two types of error probabilities. The type-I error probability (resp. type-II error probability) is the probability of deciding on $\mathcal{H}=1$ (resp. $\mathcal{H}=0$) when the original hypothesis is $\mathcal{H}=0$ (resp. $\mathcal{H}=1$).

There are two well-known approaches to the hypothesis testing setup. In one approach, the number of observed samples at the sensors is fixed and bounded by $n$ which tends to infinity. This setup is commonly referred to as \emph{fixed-length testing}. In another approach,  the sensors are allowed to get a random number of samples whose expectation is fixed and bounded by $n$. This setup is referred to as \emph{sequential testing} due to \cite{Wald}. 

The trade-off between type-I and type-II error probabilities is considered in some previous works \cite{Ahlswede,  Han, Wald, Vincent-sequential, Tara, Yury-paper, Michele, Michele2, Gunduz, Vincent, TianChen, Tuncel, SW18, W18, related-work, HA-survey}. When both error probabilities are required to decrease exponentially, the Neyman-Pearson test \cite[Thm 11.7.1]{cover} is shown to be optimal for a fixed-length test. There is a trade-off between the exponents of type-I and type-II error probabilities such that with an increase in one of the exponents, there is a decrease in the other exponent. Using sequential testing \cite{Wald}, one can resolve this trade-off and simultaneously achieve the exponents $D(Q_X\|P_X)$ and $D(P_X\|Q_X)$ for type-I and type-II error probabilities, respectively. Another regime of interest is when the type-I error probability is restricted to be smaller than some $\epsilon\in (0,1)$. In this case, Chernoff-Stein's lemma \cite[Thm 11.8.3]{cover} shows that the maximum exponent of type-II error probability is $D(P_X\|Q_X)$ which is achievable in both fixed-length and sequential testing setups.    

\begin{figure}[t]
	\centering
	\includegraphics[scale=0.4]{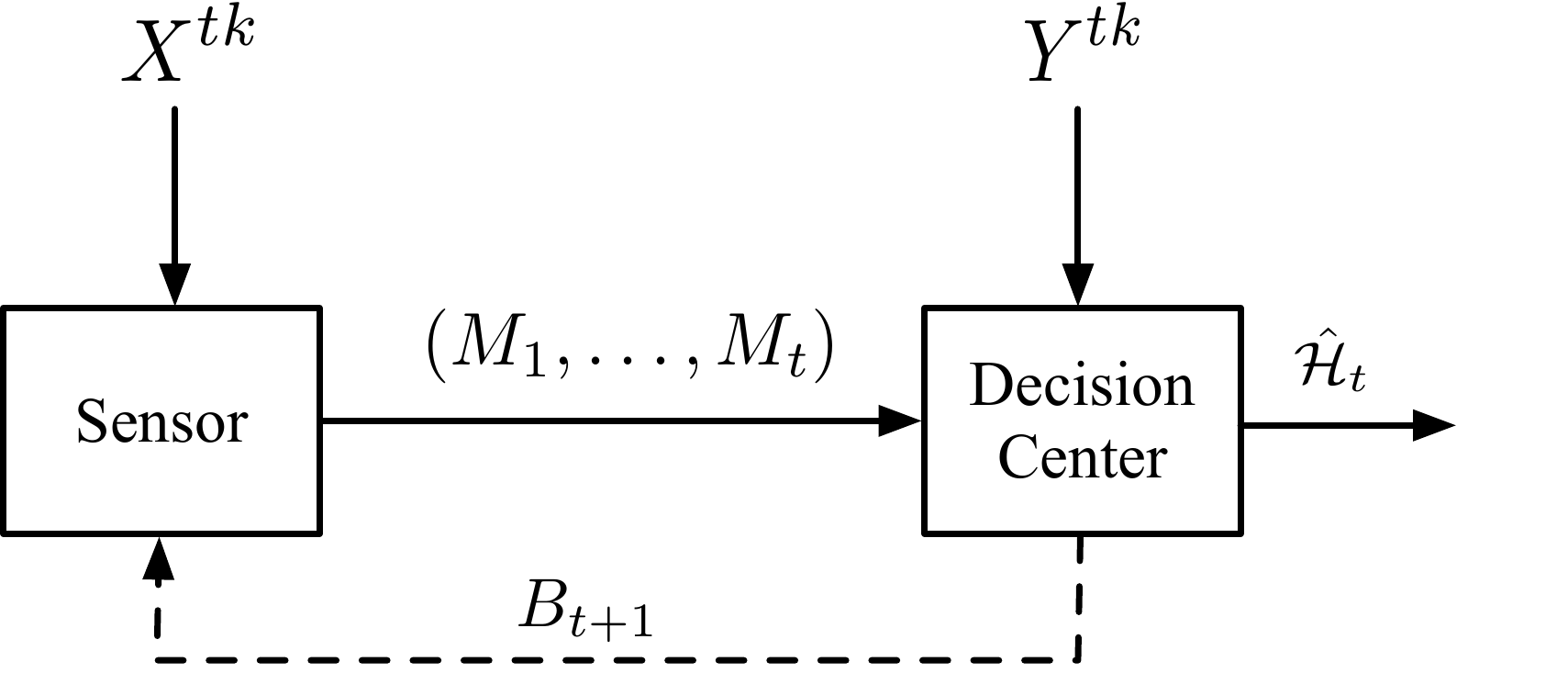}
	\caption{Distributed sequential hypothesis testing with zero-rate compression.}
	\label{fig1}
\end{figure} 

In this paper, we consider sequential testing over a simple network with a sensor and a decision center. At each time $t$, the sensor gets $k$ samples ($k>0$) of a source denoted by $X$. It then describes its observations until time $t$ over a zero-rate  noiseless link to a decision center which also has access to the samples of a source denoted by $Y$. Under the null and alternative hypotheses, each sample of the pair $(X,Y)$ is distributed according to given pmfs $P_{XY}$ and $Q_{XY}$, respectively. The decision center based on all received messages (the previous and current messages) and its observed source samples tries to decide on the hypothesis. If it makes a decision, it sends a single bit of feedback to the sensor to stop transmission which is called as \emph{stop-feedback}. However, if it needs more samples to make a decision, it sends a single bit of feedback to the sensor to request for another $k$ samples for compression and testing. The transmission continues until the decision center can finally declare a hypothesis. We are interested to find the maximum exponent of the type-II error probability under the constraint that type-I error probability does not exceed some positive $\epsilon\in (0,1)$ and when the expectation of the number of requests from the sensor does not exceed some large $n$. 

The optimal exponent of fixed-length testing over a single-sensor, single-decision center setup with a zero-rate link has been established in \cite{ShalabyPapamarcou}. Interestingly, we show that the optimal exponent of the proposed sequential testing setup coincides with that of \cite{ShalabyPapamarcou}. The proof of achievability is straightforward since the decision center can send its transmission request to the sensor $n $ times   and the compression/testing scheme  of \cite{ShalabyPapamarcou} can be employed. The main technical contribution of this paper is the proof of  the strong converse which shows that this scheme is indeed optimal for sequential testing. The proof uses Marton's blowing up lemma \cite{MartonBU}. 
\section{System Model}\label{sec:model}

Let $\set{X}$ and $\set{Y}$ be arbitrary finite alphabets and $n$ and $k$ be  positive integers. Consider the distributed sequential hypothesis testing as in Fig.~\ref{fig1}. At each time $t\in\mathbb{Z}^+$ , the sensor gets $k$ samples $X_{(t-1)k+1}^{tk}\eqdef(X_{(t-1)k+1}, \ldots, X_{tk})\in \set{X}^{ k}$ and encodes this sequence  to a zero-rate message $M_t\in \mathcal{M}_t^{(k)}\triangleq \{0,1,\ldots,|\mathcal{M}_t^{(k)}|\}$ using the function $f_t^{(k)}:\mathcal{X}^{tk}\to \mathcal{M}_t^{(k)}$ such that $M_t=f_t^{(k)}(X^{tk})$. 
The sensor sends the message $M_t$ over a noiseless zero-rate link to the decision center which also has access to side information $Y^{tk}\in\mathcal{Y}^{tk}$. Under the null hypothesis 
\begin{align}
\mathcal{H}=0\colon \quad (X,Y)\sim \; P_{XY},  \label{eqn:null}
\end{align}
whereas under the alternative hypothesis 
\begin{align}
\mathcal{H}=1\colon \quad (X,Y)\sim \; Q_{XY},\label{eqn:alt}
\end{align}
for two given pmfs $P_{XY}$ and $Q_{XY}$ where we assume the positivity constraint $Q_{XY}>0$.

At each time $t$, the decision center uses the messages $(M_1,\ldots,M_t)\in \mathcal{M}_1^{(k)}\times \ldots\times \mathcal{M}_t^{(k)}$ and the side information $Y^{tk}$ to produce an estimate of the hypothesis $\hat{\mathcal{H}}_t\in\{0,1,\star\}$ using the function \begin{align}g_t^{(k)}\colon \mathcal{M}_1^{(k)}\times\ldots \mathcal{M}_t^{(k)}\times \mathcal{Y}^{tk}\to \{0,1,\star\},\end{align} such that $\hat{\mathcal{H}}_t=g_t^{(k)}(M_1,\ldots,M_t,Y^{tk})$.

If $\hat{\mathcal{H}}_t\in\{0,1\}$, it sends a single bit of feedback $B_{t+1}=0$ to the sensor to stop transmission. If $\hat{\mathcal{H}}_t=\star$, it sends the bit $B_{t+1}=1$ to the sensor to request for more samples of the source sequence.  Define $T$ to be the stopping time of the transmission:
\begin{IEEEeqnarray}{rCl}
	T\triangleq \min\{ t\colon B_t=0 \}.
	\end{IEEEeqnarray}
Notice that $B_{t+1}$ is a function of both $X^{tk}$ and $Y^{tk}$ in which case $T$ is stopping time with respect to the filtration $\{\sigma(X^{tk},Y^{tk})\}_{t=1}^{\infty}$. Let $\mathcal{F}_T$ be the $\sigma$-algebra generated by the random variables $\{ (X^{Tk} , Y^{Tk})\}$.

The final decision function is $g_{T}^{(k)}\colon \mathcal{M}_1^{(k)}\times\ldots \mathcal{M}_T^{(k)}\times \mathcal{Y}^{T k}\to \{0,1\},$ such that $\hat{\mathcal{H}}_{T}=g_{T}^{(k)}(m_1,\ldots,m_{T},y^{T k})$.

We define  an acceptance region $\mathcal{A}_{T}\in\mathcal{F}_T$ such that:
\begin{align}
&\mathcal{A}_{T} \eqdef \Big\{(x^{Tk},y^{T k})\colon \nonumber\\&\hspace{1.1cm}  g_{T}^{(k)}(f_1^{(k)}(x^k),\ldots, f_T^{(k)}(x^{Tk}),y^{T k})=0 \Big\},
\end{align}
and a rejection region $\set{R}_{T}\in\mathcal{F}_T$ as the following:
\begin{align}
&\mathcal{R}_{T} \eqdef  \Big\{(x^{Tk},y^{T k})\colon \nonumber\\&\hspace{1.1cm}  g_{T}^{(k)}(f_1^{(k)}(x^k),\ldots, f_T^{(k)}(x_{(T-1)k+1}^{Tk}),y^{T k})=1 \Big\}.
\end{align}
Notice that $\mathcal{F}_T= \mathcal{A}_{T}\cup \mathcal{R}_{T}$.

\begin{definition}\label{def} For a given $\epsilon\in (0,1)$, we say that a type-II  exponent  $\theta\in \Reals_+$ is $\epsilon$-achievable 
	if there exists a sequence of encoding and decision functions such that the corresponding sequences of type-I  and type-II error probabilities at the decision center are respectively defined as
	\begin{align}\alpha_n\eqdef P^{Tk}_{XY}(\set{R}_{T})\quad\mbox{and}\quad\beta_n\eqdef Q_{XY}^{Tk}(\set{A}_{T}),\end{align}
	and they satisfy 
	\begin{align}
	\alpha_n\leq \epsilon \quad,\quad 	\liminf_{k\to\infty}\liminf_{n\to\infty}\frac{1}{nk}\log\frac{1}{\beta_n}\geq \theta,\label{def-exponent}
	\end{align}
	and the stopping time satisfies:
\begin{equation}
\max\{\mathbb{E}_P[T],\mathbb{E}_Q[T]\}\leq n.\label{def-time}
\end{equation}
We assume zero-rate compression in this paper, which means
\begin{IEEEeqnarray}{rCl}
	\lim_{k\to\infty} \frac{1}{k}\log|\mathcal{M}_t^{(k)}|=0,\qquad t \in \mathbb{Z}^+.\label{zero-rate-cons}
\end{IEEEeqnarray}
		The {\em optimal $\epsilon$-exponent} $\theta^*(\epsilon)$ is the supremum of all $\epsilon$-achievable exponents $\theta\in \Reals_+$.
\end{definition}
\begin{remark} We remark that the order of limits in \eqref{def-exponent} is important. First, we fix the number of samples $k \in \mathbb{Z}^+$ of the source $X$ and side information $Y$. Then, we consider the sequence of functions (decoders) $g_t^{ (k) }$ under the constraint that the stopping time satisfies \eqref{def-time}. This corresponds to a sequence of subblocks $\{(X_{(t-1)k+1}^{tk} \}_{t=1}^\infty$ of size $k$ that allow for us to make a decision confidently. Then, we take $k$ to be large and, in particular, satisfies the zero rate constraint in \eqref{zero-rate-cons}. In the traditional setup, the roles of $n$ and $k$ are merged. However, here, we separate the size $k$  of the submessages $M_t$ and the number of such submessages to make a decision. 
	\end{remark}

\section{Main Result}

The following theorem establishes the optimal exponent of the above setup. Interestingly, this exponent coincides with that of \cite{ShalabyPapamarcou} for the fixed-length testing. This implies that there is no improvement in the performance of sequential testing comparing to fixed-length setup.
\begin{theorem}\label{main-thm} Assuming that $\min_{x,y}Q_{XY}(x,y)>0$, the optimal $\epsilon$-exponent of the distributed sequential HT with zero-rate compression for all $0<\epsilon<1$ is given by
	\begin{IEEEeqnarray}{rCl}
		\theta^*(\epsilon) &= & \min_{\substack{\tilde{P}_{XY}:\\\tilde{P}_X=P_X\\\tilde{P}_Y=P_Y}}D(\tilde{P}_{XY}\|Q_{XY}).
		\end{IEEEeqnarray}
	\end{theorem}
\begin{IEEEproof}  The achievability follows from the fixed-length testing scheme of \cite{ShalabyPapamarcou}. Notice the fact that any achievable exponent for fixed-length testing is also achievable for the sequential setup since we can always take a fixed number of samples from source sequences but this may be suboptimal for sequential testing. However, as it is proved in the following Section~\ref{sec-proof}, this strategy is indeed optimal for the proposed distributed testing setup.
	\end{IEEEproof}\medskip

\subsection{Proof of Converse for Theorem~\ref{main-thm}}\label{sec-proof} \medskip

‌Before starting the proof, we present a useful lemma which will be used later.

\begin{lemma}\label{lem-main} Let $T$ be a random variable and for each $T=t$, let $P^t$ and $Q^{t}$ be  arbitrary product distributions over a set $\mathcal{Z}^t$ where $Q>0$ and $\mathcal{A}^t$ be a subset of $\mathcal{Z}^t$. Then,
	\begin{IEEEeqnarray}{rCl}
		-\mathbb{E}\left[P^T(\mathcal{A}^T)\log Q^T(\mathcal{A}^T)\right]\leq \mathbb{E}[T]D(P\|Q)+1.
	\end{IEEEeqnarray} 
\end{lemma}
\begin{IEEEproof} See Section~\ref{lem-proof-sec}.
	\end{IEEEproof}

Now, we state the proof. Fix $\epsilon \in (0,1)$ and an achievable exponent $\theta<\theta^*(\epsilon)$, a sequence of encoding and decision functions, a filtration, stopping time, acceptance and rejection regions $\mathcal{A}_{T}$, $\mathcal{R}_{T}$ such that \eqref{def-exponent}, \eqref{def-time} and \eqref{zero-rate-cons} are satisfied. Further fix a large blocklength $n$. Let $d$ be a positive integer such that \begin{align}d>\frac{\epsilon}{1-\epsilon},\label{d-const}\end{align}
and define
\begin{align}\tau\triangleq n\cdot (1+d).\label{tau-def}\end{align} 

Moreover, we define a new acceptance region $\mathcal{A}^{\text{new}}\subseteq \mathcal{A}_{T}$ such that 
\begin{IEEEeqnarray}{rCl}
	\mathcal{A}^{\text{new}} &\triangleq &\{ (x^{Tk},y^{Tk})\in\mathcal{X}^{Tk}\times \mathcal{Y}^{Tk}\colon\nonumber\\&&\hspace{3cm} (x^{Tk},y^{Tk})\in\mathcal{A}_{T},\;\; T\leq \tau \}.\nonumber\\ 
	\end{IEEEeqnarray}
Now, consider the following sets of inequalities:
\begin{IEEEeqnarray}{rCl}
	1-\epsilon&\leq& P_{XY}^{Tk}(\mathcal{A}_{T})\\
	&=& \Pr[T\leq \tau]\cdot P_{XY}^{Tk}(\mathcal{A}_{T}|T\leq \tau)\nonumber\\
	&&\hspace{0.3cm} +\Pr[T> \tau]\cdot P_{XY}^{Tk}(\mathcal{A}_{T}|T> \tau)\\
	&\leq&  P_{XY}^{Tk}(\mathcal{A}^{\text{new}})+\frac{\mathbb{E}[T]}{\tau} \\
	&\leq&  P_{XY}^{Tk}(\mathcal{A}^{\text{new}})+\frac{1}{1+d}.\label{step1}
	\end{IEEEeqnarray}
Define 
\begin{IEEEeqnarray}{rCl}\phi\triangleq 1-\epsilon-\frac{1}{1+d},\label{eps-def}\end{IEEEeqnarray}
 and notice that $0<\phi<1$ from \eqref{d-const} and  by~\eqref{step1}, we have:
\begin{IEEEeqnarray}{rCl}
	\phi \leq P_{XY}^{Tk}(\mathcal{A}^{\text{new}}).\label{step2}
	\end{IEEEeqnarray}
Now,  for $\mathsf{m}^T\eqdef (m_1,\ldots,m_T)\in\mathcal{M}_1^{(k)}\times\ldots\times\mathcal{M}_T^{(k)}
$, we define the following sets 
\begin{IEEEeqnarray}{rCl}
C_{\mathsf{m}^T}&\triangleq & \nonumber\\&&\hspace{-0.5cm}\{ x^{Tk}\in\mathcal{X}^{Tk}\colon (f_1^{(k)}(x^k),\ldots,f_T^{(k)}(x^{Tk}))=\mathsf{m}^{T} \},\nonumber\\\\
F_{\mathsf{m}^T}&\triangleq & \{ y^{Tk}\in\mathcal{Y}^{Tk}\colon g_T^{(k)}(\mathsf{m}^T,y^{Tk})=0 \}.
\end{IEEEeqnarray}
The sets $C_{\mathsf{m}^T}$ (and $F_{\mathsf{m}^T}$) for different $\mathsf{m}^T$ are disjoint. That is, for each two message sequences $\mathsf{m}^T$, $\mathsf{m}'^T$ such that $\mathsf{m}^T\neq \mathsf{m}'^T$, we have:
\begin{IEEEeqnarray}{rCl}
	C_{\mathsf{m}^T}\cap C_{\mathsf{m}'^T}=\emptyset,\qquad F_{\mathsf{m}^T}\cap F_{\mathsf{m}'^T}=\emptyset.
\end{IEEEeqnarray}
Given the above sets, we can write 
\begin{IEEEeqnarray}{rCl}
	\mathcal{A}^{\text{new}} &=& \bigcup_{\mathsf{m}^T}C_{\mathsf{m}^T}\times F_{\mathsf{m}^T}, \qquad T\leq \tau.\label{step3}
	\end{IEEEeqnarray}
Considering \eqref{step2} and \eqref{step3}, there exists a message $\mathsf{m}^*$ such that
\begin{IEEEeqnarray}{rCl}
P_{XY}^{Tk}(C_{\mathsf{m}^*}\times F_{\mathsf{m}^*})\geq \frac{\phi}{\prod_{t=1}^T|\mathcal{M}_t^{(k)}|}.\label{step4}
	\end{IEEEeqnarray}
Let $C\triangleq C_{\mathsf{m}^*}$, $F\triangleq F_{\mathsf{m}^*}$ and 
\begin{IEEEeqnarray}{rCl}
		\delta_{k,\tau }&\triangleq &-\frac{1}{\tau k}\log \phi+\frac{1}{\tau k}\sum_{t=1}^T\log |\mathcal{M}_t^{(k)}|,
	\end{IEEEeqnarray}
where $\delta_{k,\tau}\to 0$ as $k,\tau\to\infty$ by \eqref{zero-rate-cons}, \eqref{tau-def}, \eqref{eps-def} and considering the fact that $T\leq \tau$.

 Thus, we can re-write \eqref{step4} as follows:
\begin{IEEEeqnarray}{rCl}
	P_{XY}^{Tk}(C\times F)\geq 2^{-\tau k\delta_{k,\tau }}.\label{step5}
	\end{IEEEeqnarray}
We define $\mathcal{A}\triangleq C\times F$ and the above inequality can be equivalently written as
\begin{IEEEeqnarray}{rCl}
	P_{XY}^{Tk}(\mathcal{A})\geq 2^{-\tau k\delta_{k,\tau }}.
	\end{IEEEeqnarray}
We then expand the region $\mathcal{A}$ to a subset of $\mathcal{X}^{\tau k}\times \mathcal{Y}^{\tau k}$ so that all sequences to be of the same length $\tau $:
\begin{IEEEeqnarray}{rCl}
	\mathcal{A}^{\text{exp}}&\eqdef& \{ (x^{\tau k},y^{\tau k})\colon \nonumber\\&&\hspace{0.5cm}\exists (\tilde{x}^{T k},\tilde{y}^{T k})\in \mathcal{A}\;\text{and}\;(\bar{x}^{(\tau-T)k},\bar{y}^{(\tau-T)k})\colon \nonumber\\&&\hspace{0.5cm}(x^{\tau k },y^{\tau k})=(\tilde{x}^{Tk},\bar{x}^{(\tau-T)k},\tilde{y}^{Tk},\bar{y}^{(\tau-T)k}) \}.\nonumber\\
	\end{IEEEeqnarray} 
 Notice that 
\begin{IEEEeqnarray}{rCl}
	P_{XY}^{\tau k}(\mathcal{A}^{\text{exp}})=P_{XY}^{Tk}(\mathcal{A})\geq 2^{-\tau k \delta_{k,\tau}}.
\end{IEEEeqnarray}
We decompose the set $\mathcal{A}^{\text{exp}}$ into two sets $C^{\text{exp}}\subseteq\mathcal{X}^{\tau k}$ and $F^{\text{exp}}\subseteq\mathcal{Y}^{\tau k }$ such that $\mathcal{A}^{\text{exp}}=C^{\text{exp}}\times F^{\text{exp}}$.
The above inequality implies that 
\begin{IEEEeqnarray}{rCl}
	P_{X}^{\tau k}(C^{\text{exp}})\geq 2^{-\tau k \delta_{k,\tau }},\qquad P_{Y}^{\tau k}( F^{\text{exp}})\geq 2^{-\tau k \delta_{k,\tau }}.\label{step7}
	\end{IEEEeqnarray}
Now, define \begin{align}\nu\eqdef \tau k,\end{align}
and let $\{\ell_{\nu} \}$ be any sequence that satisfies 
\begin{IEEEeqnarray}{rCl}
	&&	\lim_{\nu \to\infty}\frac{\ell_{\nu}}{\sqrt{\nu\log \nu}}=\infty,\\
	&& \lim_{\nu\to\infty}\frac{\ell_{\nu}}{\nu}=0	
\end{IEEEeqnarray}
Using the blowing-up lemma \cite[Remark on p. 446]{MartonBU}, we get:
\begin{IEEEeqnarray}{rCl}
	P_{X}^{\nu }(C^{\text{exp-bl},\ell_{\nu}})\geq 1-\xi_{\nu},\;\;\; P_{Y}^{\nu }(F^{\text{exp-bl},\ell_{\nu}})\geq 1-\xi_{\nu},\label{blowup-con}
	\end{IEEEeqnarray}
where we define $\ell_{\nu}$-blown up sets of $C^{\text{exp}}$ and $F^{\text{exp}}$ as follows:
\begin{IEEEeqnarray}{rCl}	
	C^{\text{exp-bl},\ell_{\nu}}&\triangleq & \{\tilde{x}^{\nu }\colon \exists x^{\nu }\in C^{\text{exp}}\;\text{s.t.}\; d_{\text{H}}(\tilde{x}^{\nu },x^{\nu })\leq \ell_{\nu} \},\\
F^{\text{exp-bl},\ell_{\nu}}&\triangleq & \{\tilde{y}^{\nu}\colon \exists y^{\nu }\in F^{\text{exp}}\;\text{s.t.}\; d_{\text{H}}(\tilde{y}^{\nu},y^{\nu })\leq \ell_{\nu} \},
	\end{IEEEeqnarray}
 and $\xi_{\nu}\triangleq \frac{\sqrt{ \nu\log \nu}}{\ell_{\nu}}$.
 
 Next, we introduce a distribution $\tilde{P}_{XY}$ that satisifes the marginal constraints 
 \begin{IEEEeqnarray}{rCl}
 \tilde{P}_X=P_X,\qquad \tilde{P}_Y=P_Y. \label{marginals}
 \end{IEEEeqnarray}
  For this distribution, we have:
 \begin{IEEEeqnarray}{rCl}
 \tilde{P}_{XY}^{\nu }(C^{\text{exp-bl},\ell_{\nu}}\times F^{\text{exp-bl},\ell_{\nu}}) &\geq & 	\tilde{P}_{X}^{\nu }(C^{\text{exp-bl},\ell_{\nu}}) +	\tilde{P}_{Y}^{\nu }( F^{\text{exp-bl},\ell_{\nu}}) \nonumber\\&&\hspace{0.5cm}-1\\
 &\geq & 1-2\xi_{\nu},\label{step10}
 	\end{IEEEeqnarray}
 where the first inequality follows from the property $\Pr(A\cap B)\geq \Pr(A)+\Pr(B)-1$ and the second inequality follows from~\eqref{blowup-con}. 
 
We define $\mathcal{A}^{\text{exp-bl},2\ell_{\nu}}\triangleq C^{\text{exp-bl},\ell_{\nu}}\times F^{\text{exp-bl},\ell_{\nu}}$ and observe that it is the $2\ell_{\nu}$-blown-up of the set $\mathcal{A}^{\text{exp}}$. It is also the expanded region of the $2\ell_{\nu}$-blown up of the set $\mathcal{A}$ which we denote by $\mathcal{A}^{\text{bl},2\ell_{\nu}}$. 
 Thus we have:
\begin{IEEEeqnarray}{rCl}
\tilde{P}_{XY}^{Tk}(\mathcal{A}^{\text{bl},2\ell_{\nu}})&=&\tilde{P}_{XY}^{\nu }(\mathcal{A}^{\text{exp-bl},2\ell_{\nu}})\\&=&\tilde{P}_{XY}^{\nu }(C^{\text{exp-bl},\ell_{\nu}}\times F^{\text{exp-bl},\ell_{\nu}}) 	\\
&\geq &1-2\xi_{\nu},\label{step18}
	\end{IEEEeqnarray}
where the last inequality follows from \eqref{step10}.
Now, we consider the following sets of inequalities:
\begin{IEEEeqnarray}{rCl}
	Q_{XY}^{Tk}(\mathcal{A}^{\text{bl},2\ell_{\nu}}) &\leq & Q_{XY}^{Tk}(\mathcal{A})\cdot K_{\nu}^{\ell_{\nu}}\label{step9}\\
	&\leq & \beta_n\cdot K_{\nu}^{\ell_{\nu}},\label{step8}
	\end{IEEEeqnarray}
where 
\begin{IEEEeqnarray}{rCl}
	K_{\nu} \triangleq \frac{\nu }{q^2\ell_{\nu}}|\mathcal{X}||\mathcal{Y}|,
	\end{IEEEeqnarray}
and 
\begin{IEEEeqnarray}{rCl}
	q\eqdef \min_{x\in\mathcal{X},\;y\in\mathcal{Y}} \;Q_{XY}(x,y).
	\end{IEEEeqnarray}
Here, \eqref{step9} follows from \cite[proof of Lemma 5.1]{Csiszarbook}, and \eqref{step8} follows because $\mathcal{A}\subseteq \mathcal{A}_{T}$. Here is where we use the fact that $q>0$ so $K_{\nu}$ is finite.

 Now, notice that the second inequality of \eqref{def-exponent} together with \eqref{step8} yields the following:
 \begin{IEEEeqnarray}{rCl}
 	\theta
 	&\leq &-\frac{1}{nk}\log \beta_n+\psi_{n,k}\nonumber\\&\leq& -\frac{1}{nk}\log 	Q_{XY}^{Tk}(\mathcal{A}^{\text{bl},2\ell_{\nu}})+\frac{\ell_{\nu}}{kn}\log K_{\nu}+\psi_{n,k},
 \end{IEEEeqnarray}
where $\psi_{n,k}\to 0$ as $n$ and $k$ tend to infinity.
The above inequality holds for any realization of $T$ so it is also satisfied when averaged over $T$. Thus, we get the following:
\begin{IEEEeqnarray}{rCl} 	\theta&\leq& -\frac{1}{nk}\mathbb{E}\left[\log Q_{XY}^{Tk}(\mathcal{A}^{\text{bl},2\ell_{\nu}})\right]+\frac{\ell_{\nu}}{kn}\log K_{\nu}+\psi_{n,k}.
	\end{IEEEeqnarray}
Thus, the constraint is also satisfied when averaged over  $T$. We continue with the following set of inequalities:
\begin{IEEEeqnarray}{rCl}
&& \hspace{-0.7cm}-\frac{1}{nk}\mathbb{E}\left[\log 	Q_{XY}^{Tk}(\mathcal{A}^{\text{bl},2\ell_{\nu}})\right]\\[1ex]
	&\leq& -\frac{1}{nk(1-2\xi_{\nu})}\mathbb{E}\left[\tilde{P}_{XY}^{Tk}(\mathcal{A}^{\text{bl},2\ell_{\nu}})\log 	Q_{XY}^{Tk}(\mathcal{A}^{\text{bl},2\ell_{\nu}})\right]\label{step17}\\[1ex]
	&=& \frac{1}{nk(1-2\xi_{\nu})}(\mathbb{E}[T] D(\tilde{P}_{XY}\|Q_{XY})+1)\label{step12}\\[1ex]
	&\leq & \frac{1}{nk(1-2\xi_{\nu})}(nD(\tilde{P}_{XY}\|Q_{XY})+1),\label{step13}
	\end{IEEEeqnarray}
where \eqref{step17}   follows from inequality \eqref{step18}, \eqref{step12} follows from Lemma~\ref{lem-main} and \eqref{step13} follows from~\eqref{def-time}. \medskip

Finally, letting $n,k\to \infty$, we have $\psi_{n,k}\to 0$, $\xi_{\nu}\to 0$ and $\frac{\ell_{\nu}}{kn}\log K_{\nu}\to 0$. Also, recall that  distribution $\tilde{P}_{XY}$ satisfies the marginal constraints in~\eqref{marginals}. Thus, we get:
\begin{IEEEeqnarray}{rCl}
\lim_{k\to\infty}\liminf_{n\to\infty}\;-\frac{1}{nk}\log \beta_n\leq \min_{\substack{\tilde{P}_{XY}:\\\tilde{P}_X=P_X\\\tilde{P}_Y=P_Y}}D(\tilde{P}_{XY}\|Q_{XY}).\nonumber\\
	\end{IEEEeqnarray}
This completes the proof.

\section{Proof of Lemma~\ref{lem-main}}\label{lem-proof-sec}

 First, we show that 
	\begin{IEEEeqnarray}{rCl}
			\mathbb{E}\left[D(P^T\|Q^T)\right] &=& \mathbb{E}[T] D(P\|Q).\label{step14}
		\end{IEEEeqnarray}
We follow similar steps to \cite[Proof on pp. 171]{Yury-book} where we introduce the following random variable:
\begin{IEEEeqnarray}{rCl}
	S_t \eqdef \sum_{i=1}^t D(P_i\|Q_i)-tD(P\|Q),
\end{IEEEeqnarray}
 for $t\geq 1$ and $S_0=0$.
Clearly, $S_t$ is a martingale w.r.t. $\mathcal{F}_t$. Thus, $\tilde{S}_t\eqdef S_{\min(T,t)}$ is also a martingale. Therefore, we have:
\begin{IEEEeqnarray}{rCl}
	\mathbb{E}[\tilde{S}_t]=\mathbb{E}[\tilde{S}_0]=0.
\end{IEEEeqnarray}
This yields:
\begin{IEEEeqnarray}{rCl}
	\mathbb{E}\left[\sum_{i=1}^{\min(T,t)} D(P_i\|Q_i)\right]=\mathbb{E}[\min(T,t)]D(P\|Q).
\end{IEEEeqnarray}
Since we have assumed that $Q>0$, then $\sum_{i=1}^tD(P_i\|Q_i)\leq t c$ for some positive $c$. Thus, $\sum_{i=1}^{\min(T,t)}D(P_i\|Q_i)\leq t c$ which implies that the following collection \begin{IEEEeqnarray}{rCl}\left\{\sum_{i=1}^{\min(T,t)}D(P_i\|Q_i),\;\;t\geq 0\right\}\end{IEEEeqnarray} is uniformly integrable.
Therefore, we can take $t\to\infty$ and interchange the limit and expectation to get to~\eqref{step14}.	

Next, consider the following set of inequalities:
\begin{IEEEeqnarray}{rCl}
	&&\hspace{-1cm}\mathbb{E}[T]D(P\|Q)\nonumber\\ &=& \mathbb{E}\left[D(P^T\|Q^T)\right]\\
	&\geq & \sum_t P_T(t)\Bigg[P^t(\mathcal{A}^t)\log \frac{P^t(\mathcal{A}^t)}{Q^t(\mathcal{A}^t)}\nonumber\\
	&&\hspace{1.3cm}+(1-P^t(\mathcal{A}^t))\log \frac{(1-P^t(\mathcal{A}^t))}{(1-Q^t(\mathcal{A}^t))}\Bigg]\nonumber\\\label{step15}\\
	&=& \sum_t P_T(t)\Bigg[-H_{\text{b}}(P^t(\mathcal{A}^t))\nonumber\\&&\hspace{1.5cm}-P^t(\mathcal{A}^t)\log Q^t(\mathcal{A}^t)\nonumber\\&&\hspace{1.5cm}-(1-P^t(\mathcal{A}^t))\log (1-Q^t(\mathcal{A}^t))\Bigg]\nonumber\\\\
	&\geq & -1-\mathbb{E}\left[P^T(\mathcal{A}^T)\log Q^T(\mathcal{A}^T)\right], \label{step16}
	\end{IEEEeqnarray}
where~\eqref{step15} follows from the data processing inequality for KL-divergence; \eqref{step16} follows from upper bounding $H_{\text{b}}(P^t(\mathcal{A}^t))$ by $1$ and $(1-P^t(\mathcal{A}^t))\log (1-Q^t(\mathcal{A}^t))$ by $0$. Finally, rearranging terms in \eqref{step16} yields the desired inequality.

\section{Conclusion}

In this paper, we considered sequential testing over a single-sensor, a single-decision center setup which communicate over a zero-rate noiseless link. We established the optimal exponent of type-II error probability under a constrained type-I error probability and when the expected number of transmission times is smaller than $n$ which tends to infinity. Interestingly, this exponent coincides with that of fixed-length testing in \cite{ShalabyPapamarcou}.  
	
\section{Acknowledgements} 
The authors would like to thank Prof. Mich\`ele Wigger for her comments. 

V.~Y.~F.~Tan is supported by a Singapore National Research Foundation (NRF) Fellowship under grant number R-263-000-D02-281.

\bibliographystyle{IEEEtran}

\bibliography{references}

\end{document}